\documentclass[twocolumn,english,aps,prl,showpacs]{revtex4}
\usepackage[T1]{fontenc}
\usepackage{amsmath,graphicx,amssymb,epsfig,babel,dsfont}

\renewcommand{\Re}{{\rm Re}}
\renewcommand{\Im}{{\rm Im}}
\newcommand{\Tr}{{\rm Tr}}
\newcommand{\rd}{{\rm d}}

\newcommand{\ri}{{\rm i}}

\begin{document}

\title{Anomalous photon thermal Hall effect}

\author{A. Ott}
\affiliation{Institut f\"{u}r Physik, Carl von Ossietzky Universit\"{a}t,
D-26111 Oldenburg, Germany.}

\author{S.-A. Biehs}
\email{s.age.biehs@uni-oldenburg.de}
\affiliation{Institut f\"{u}r Physik, Carl von Ossietzky Universit\"{a}t,
D-26111 Oldenburg, Germany.}

\author{P. Ben-Abdallah}
\email{pba@institutoptique.fr} 
\affiliation{Laboratoire Charles Fabry, UMR 8501, Institut d'Optique, CNRS, Universit\'{e} Paris-Saclay, 2 Avenue Augustin Fresnel, 91127 Palaiseau Cedex, France.}

\date{\today}

\begin{abstract}
We predict an anomalous thermal Hall effect (ATHE) mediated by photons in networks of Weyl semi-metals. Contrary to the photon thermal Hall effect in magneto-optical systems which requires the application of an external  magnetic field the ATHE in a Weyl semi-metals network is an intrinsic property of these systems. Since the  Weyl semi-metals can exhibit a strong nonreciprocal response in the infrared over a broad spectral range the magnitude of thermal Hall flux in these systems can be relatively large compared to the primary flux. This ATHE  paves the way for a directional control of heat flux  by  localy tuning the magnitude of temperature field without changing the direction of temperature gradient. 
\end{abstract}

\maketitle

The classical Hall effect~\cite{Hall} results in the appearance of a transverse electric current inside a conductor under the action of an external magnetic field applied in the direction orthogonal to the primary bias voltage. This effect stems from the Lorentz force which acts transversally on the electric charges in motion through the magnetic field curving so their trajectories. Very shortly after this discovery, a thermal analog of this effect has been observed by Righi and Leduc~\cite{Leduc} when a temperature gradient is applied throughout an electric conductor. As for the classical Hall effect, this effect is intrinsically related to the presence of free electric charges. Hence, one would not expect a thermal Hall effect with neutral particles or quasiparticles. Nevertheless, during the last decade researchers have highlighted such an effect in non-conducting materials due to phonons~\cite{Strohm,Inyushkin}, magnons (spin waves)~\cite{Fujimoto,Katsura,Onose} and even photons~\cite{PBA_PRL2016,Ott_JPE} in non-reciprocal many-body systems due to different mechanisms of local broken symmetry induced by application of an external magnetic field. Beside these 'normal' Hall effects, anomalous effects~\cite{AHE_RMP,Karplus} have been predicted in solids such as ferromagnets without external field application. In these media an intrinsic mechanism (a Berry curvature acting as a fictitious field on electrons,  a skew scattering that is an asymmetric impurity scattering or a spin-orbit coupling) is responsible for the  local symmetry breaking which gives rise to a Hall current. More recently thermal analogs of this effect, also called anomalous thermal Hall effect~\cite{Ferreiros,Sugii,Huang2} have been measured in these media. 

\begin{figure}
\centering
\includegraphics[angle=0,scale=0.3]{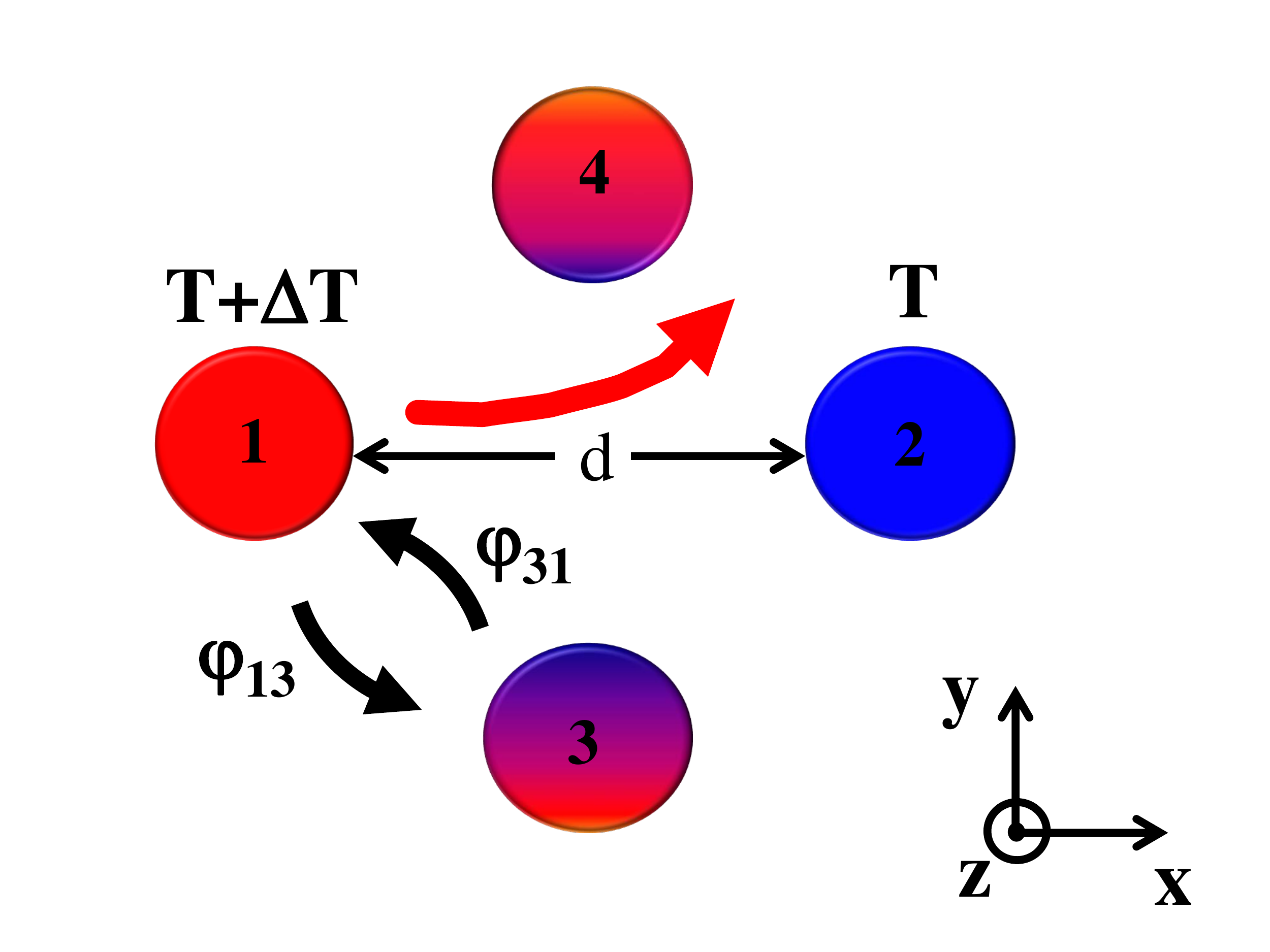}
	\caption{Four terminal junction with particles made of WSM forming a square with $C_4$ symmetry. To measure the photon thermal Hall flux a temperature gradient $\Delta T$ is applied between particles $1$ and $2$ along the $\mathbf{x}$-axis and the Hall flux is evaluated by measuring the temperature difference $T_3^{({\rm st})}-T_4^{({\rm st})}$ in the transveral direction along the $\mathbf{y}$-axis in steady state.   In this figure, the heat flux $\varphi_{13}$ exchanged from the first to the third particle is different than the flux $\varphi_{31}$ transferred in the opposite direction.\label{Fig:Sketch}} 
\end{figure} 

In this paper we predict that many-body interactions mediated by thermal photons in Weyl semi-metal (WSM) networks as depicted in Fig.~\ref{Fig:Sketch} can lead to an anomalous Hall flux. WSMs~\cite{Huang,Xu,Lv} are materials where valence and conduction bands cross in single points. Some of these media can exhibit, because of their unique topologically nontrivial electronic states~\cite{Armitage,Yan}, a strong nonreciprocal optical response~\cite{Kotov1,Kotov2}.

\begin{figure}
  \centering
  \includegraphics[angle=0,scale=0.45,angle=0]{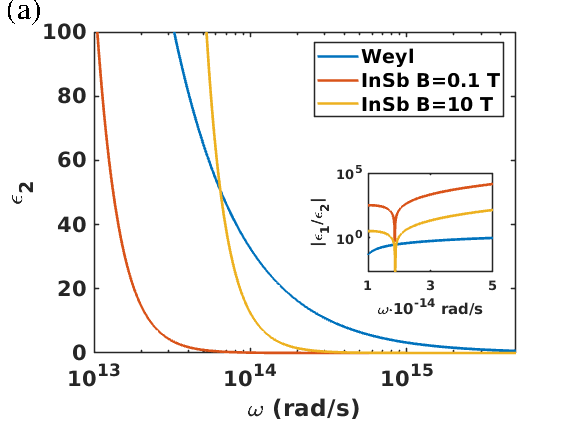}
 \includegraphics[angle=0,scale=0.45,angle=0]{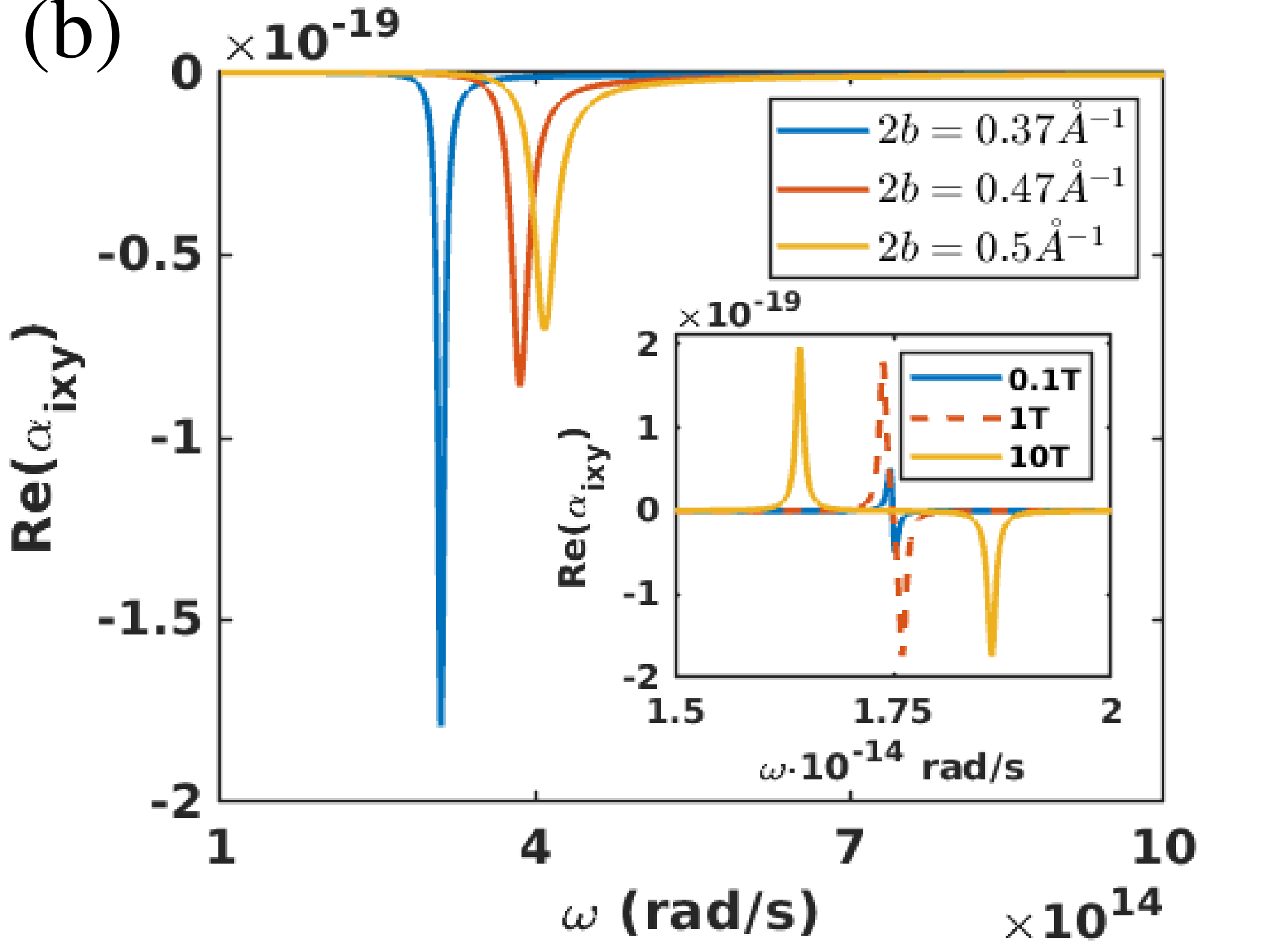}
	\caption{(a) The non-diagonal element of the permittivity tensor of the WSM and InSb (parameters are given in Ref.~\cite{SM}) with an applied magnetic field in z-direction with amplitude $B = 0.1, 10\,{\rm T}$ in the infrared range at T=300 K. Inset: $|\epsilon_1| / |\epsilon_2|$ for the same materials.
	(b) The non-diagonal element $\Re({\alpha_{xy}})$ of $(\boldsymbol{\alpha}_i - \boldsymbol{\alpha}_i^\dagger)/2 \ri$ entering the transmission coefficient in Eq.~(\ref{Eq:TransmissionCoefficient}) for WSM with $2b = 0.37, 0.47, 0.5 \,\mathring{A}^{-1}$. Inset: $\Re({\alpha_{xy}})$ for InSb (parameters are given in Ref.~\cite{SM}) with an applied magnetic field in z-direction with amplitude $B = 0.1, 1, 10\,{\rm T}$ in the infrared range at T=300 K.}
  \label{Fig:EpsilonWeyl}
\end{figure}

To investigate this effect we consider the system sketched in Fig.~1. Identical spherical particles of radius $r$ of  WSMs are arranged in a four-terminal junction. The permittivity tensor of these particles takes the following form~\cite{Kotov1,Kotov2}
\begin{equation}
  \boldsymbol{\varepsilon}=\left(\begin{array}{ccc}
  \varepsilon_{1} & -i\varepsilon_{2} & 0\\
  i\varepsilon_{2} & \varepsilon_{1} & 0\\
  0 & 0 & \varepsilon_{1}
  \end{array}\right)
\label{Eq:permittivity}
\end{equation}
with
\begin{equation}
  \varepsilon_{1}=\varepsilon_{b}+i\frac{\sigma}{\Omega}\label{Eq:permittivity1}
\end{equation}
and
\begin{equation}
  \varepsilon_{2}=\frac{b e^2}{2\pi^2\hbar\omega}\label{Eq:permittivity2}.
\end{equation}
Here 
\begin{equation}
\begin{split}
	\sigma &= \frac{r_s g}{6} \Omega G(\Omega/2)+i\frac{r_s g \Omega}{6\pi}\bigg[\frac{4}{\Omega^2}\biggl(1+\frac{\pi^2}{3}\biggl(\frac{k_B T}{E_F}\biggr)^2\biggr) \\
	       &\qquad +8\int_{0}^{\xi_c}\frac{G(\xi)-G(\Omega/2)}{\Omega^2-4\xi^2}\xi d\xi\biggr]
\label{Eq:sigma}
\end{split}
\end{equation}
where $v_F$ is the Fermi velocity,  $\Omega$ is the complex frequency normalized by the chemical potential, $g$ is the number of Weyl points, $r_s=e^2/4\pi\varepsilon_0\hbar v_F$ is the effective fine structure constant, $e$ being the electron charge, and  $\xi_c=E_c/E_F$ where Ec is the cutoff energy (see details in Refs.~\cite{Kotov1,Kotov2,SM}). If not explicitely stated we use for the WSM the parameter set $v_F = 10^6\,{\rm m/s}, g = 2, 2 b = 0.47\,\mathring{A}^{-1}$, $\xi_c = 3$, and $\tau = 10^{-12}\,{\rm s}$. Note that the permittivity tensor is non-reciprocal (i.e.\ $\boldsymbol{\varepsilon} \neq \boldsymbol{\varepsilon}^t$) so that for WSMs we intrinsically have such effects like a persistent heat currents~\cite{OttEtAl2018,zhufan,zhufan2,Silveirinha}, persistent angular momentum and spin~\cite{OttEtAl2018,Silveirinha,Zubin2019} as found for non-reciprocal magneto-optical materials and as we will show the presence of an ATHE. Contrary to usual magneto-optical materials the non-reciprocity of WSMs is strong because the separations of the Weyl nodes in momentum space $2 b$ can have relatively large values with compounds like Eu$_2$IrO$_7$, Co$_3$S$_2$Sn$_2$, and Co$_3$S$_2$Se$_2$~\cite{Kotov1}. This can be seen by comparison, for instance, with the non-diagonal term of permittivity tensor of usual magneto-optical material such as  Indium Antimonide (InSb) under a strong magnetic field $\mathbf{B} = B \mathbf{e}_z$ applied  in z direction (see Fig.~\ref{Fig:EpsilonWeyl}(a)). In this case the off-diagonal term is given by~\cite{SM}
\begin{equation}
	\epsilon_2^{\rm InSb} = \frac{\epsilon_\infty \omega_p^2 \omega_c}{\omega \bigl((\omega+ \ri\gamma)^2 - \omega_c^2\bigr)}
\end{equation}
where $\omega_p$ and $\omega_c = e B /m^*$ are the plasma and cyclotron frequency, $\gamma$ is a phenomenological damping constant,  $m^*$ the effective electron mass, and $\epsilon_\infty$ is a material specific constant factor. This term clearly depends on the magnetic field strength $B$ via $\omega_c$ and as can be seen in Fig.~\ref{Fig:EpsilonWeyl}(a) even when applying a magnetic field of $10\,{\rm T}$ we see that the non-reciprocity quantified by the ratio $|\epsilon_2|/|\epsilon_1|$ is much smaller for InSb than for WSMs over a broad spectral range in the infrared. Hence, the WSMs can be seen as a magneto-optical material under the action of an extremely large magnetic field.  

Using the Landauer formalism for N-body systems~\cite{PBA_PRL2016,PBAEtAl2011,Riccardo,Ivan_PRL2017,zhufan,zhufan2,Cuevas,Cuevas2} the non-equilibrium heat flux exchanged from the $i^{th}$ to the $j^{th}$ particle in the network reads 
\begin{equation}
	\varphi_{ij}=\int_{0}^{\infty}\frac{\rd\omega}{2\pi}\,[\Theta(\omega,T_{i})-\Theta(\omega,T_{j})] \mathcal{T}_{ij}\label{Eq:InterpartHeatFlux},
\end{equation}
where $\Theta(\omega,T)={\hbar\omega}/[{e^{\frac{\hbar\omega}{k_B T}}-1}]$ is the mean energy of a harmonic oscillator in
thermal equilibrium at temperature $T$ and $\mathcal{T}_{ij}(\omega)$ denotes the transmission coefficient, at the frequency $\omega$, between the particles $i$ and $j$. In the dipolar approximation the transmission coefficient reads~\cite{Cuevas,Cuevas2}
\begin{equation}
	\mathcal{T}_{ij}(\omega) = \frac{4}{3} k_0^4 \Im \Tr\biggl[\boldsymbol{\alpha}_j\mathds{G}_{ji}\frac{\boldsymbol{\alpha}_i-\boldsymbol{\alpha}_i^\dagger}{2 \ri}\mathds{G}_{ji}^{\dagger}\biggr],
\label{Eq:TransmissionCoefficient}
\end{equation}
where $k_0 = \omega/c$ with the light velocity $c$ and $\mathds{G}_{ij}$ denotes the dyadic Green tensor between the $i^{th}$ and the $j^{th}$ particle in the N-dipole system~\cite{Purcell} and $\boldsymbol\alpha_i$ is the polarizability tensor of the $i^{th}$ particle which reads for anisotropic particles in vacuum~\cite{Albaladejo}
\begin{equation}
   \boldsymbol{\alpha}_{i}(\omega)= \biggl( \boldsymbol{1} - i\frac{k^3_0}{6\pi} \boldsymbol{\alpha_0}_{i}\biggr)^{-1} \boldsymbol{\alpha_0}_{i}
\label{Eq:Polarizability},
\end{equation}
where $\boldsymbol{\alpha_0}_{i}$ denotes the quasi-static or undressed polarizability of the $i^{th}$ particle. For anisotropic spheres embedded in vacuum it reads~\cite{LakhtakiaEtAl1991}
\begin{equation}
  \boldsymbol{\alpha_0}_{i}(\omega)=4\pi r^3(\boldsymbol{\varepsilon} - \mathds{1})(\boldsymbol{\varepsilon} + 2 \mathds{1})^{-1}
  \label{Eq:Polarizability2}.
\end{equation}
Due to the non-reciprocity of the permittivity the polarizability tensor is non-reciprocal as well, i.e.\ $\boldsymbol{\alpha}_{0i} \neq \boldsymbol{\alpha}_{0i}^t$ and in particular ${\alpha_{0i}}_{xy} = - {\alpha_{0i}}_{yx}$. Hence, the skew hermitian part of the polarizability $(\boldsymbol{\alpha}_i - \boldsymbol{\alpha}_i^\dagger)/2 \ri$  entering in the transmission coefficient has the diagonal elements $\Im({\alpha_i}_{\nu\nu})$ ($\nu = x,y,z$) and the off-diagonal elements $\Re({{\alpha_i}_{xy}})$ and $\Re({{\alpha_i}_{yx}}) =- \Re({{\alpha_i}_{xy}})$, respectively. In Fig.~\ref{Fig:EpsilonWeyl}(b) we show $\Re({\alpha_i}_{xy})$ which is responsible for the non-reciprocity and the thermal Hall effect. The two resonances in $\Re({\alpha_i}_{xy})$ as  seen for InSb in Fig.~\ref{Fig:EpsilonWeyl}(b) correspond to the dipolar resonances $\omega_{m = \pm 1}$ of the nanoparticle, whereas the resonance $\omega_{m = 0}$ can also be observed for the diagonal elements of the polarizability. Here $m$ represents the topological charge of particles~\cite{OttEtAl2018}. For the WSM the high-frequency resonances at $\omega_{m = -1}$ dominate $\Re({\alpha_i}_{xy})$ for our choice of parameters so that only these resonances can be seen in Fig.~\ref{Fig:EpsilonWeyl}(b). The magnitude of this resonance of the WSM is for the different values of $2b$ comparable to the corresponding resonances of InSb with a magnetic field between 0.1 and 1T. This might be astonishing because we have seen before that the non-diagonal elements of the permittivity tensor of the WSM are much stronger than those of InSb with a magnetic field of 1T to 10T. But the difference is due to the fact that for the WSM the resonance of $\varepsilon_2$ is at $\omega = 0$ and that of InSb is at $\omega_c$. By increasing the magnetic field the $m = \pm 1$ resonances frequency split and are separated by $\omega_{m = -1} - \omega_{m = +1} \approx \omega_c$ while their magnitude increases as can be observed in Fig.~\ref{Fig:EpsilonWeyl}(b). For the WSM the impact of the large non-diagonal elements of the permittivity tensor is that the splitting of the two resonances is much larger than for InSb, but the magnitude of the resonances does not get so large.

\begin{figure}
  \centering
  \includegraphics[angle=0,scale=0.45,angle=0]{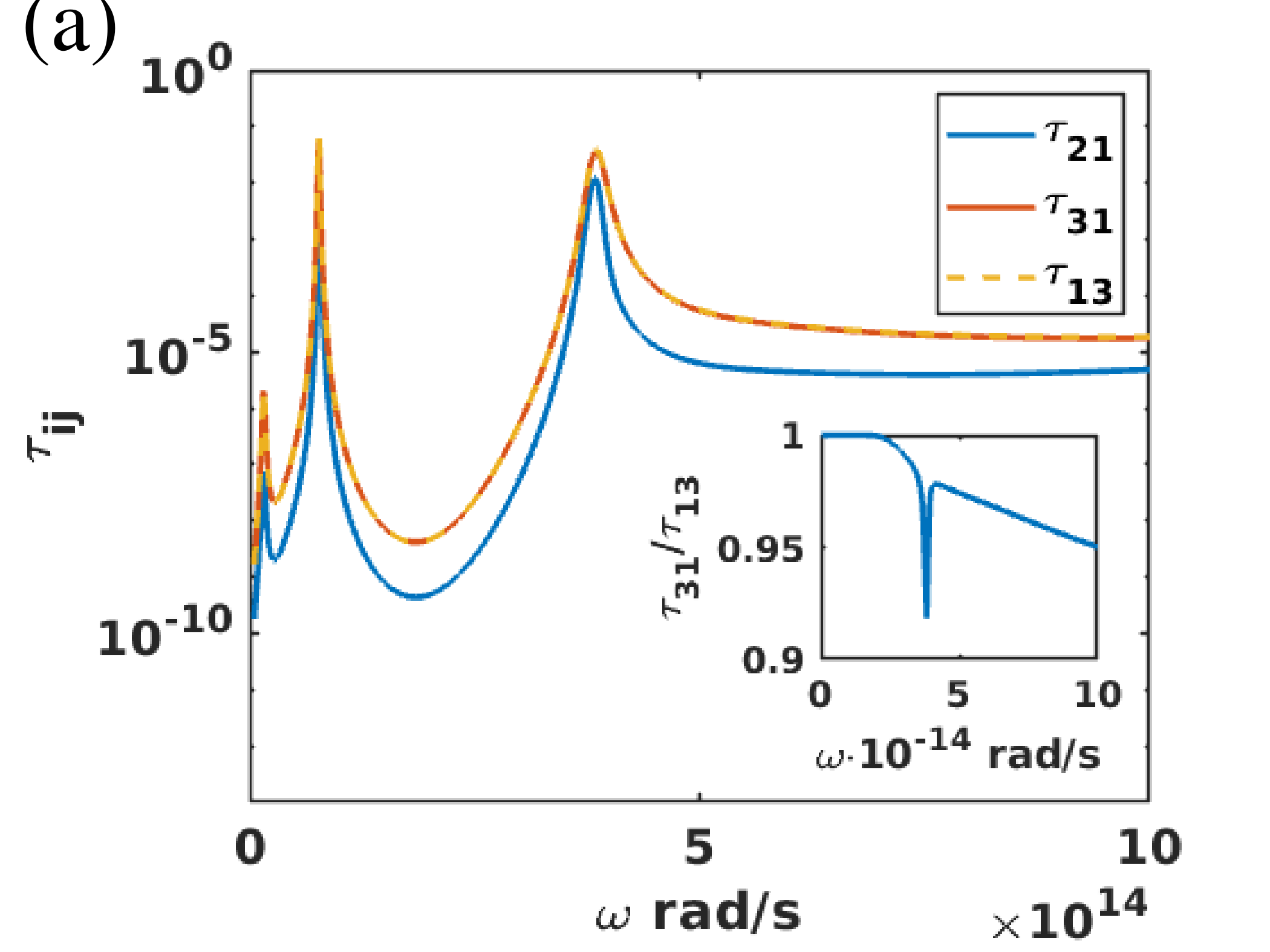}
  \includegraphics[angle=0,scale=0.45,angle=0]{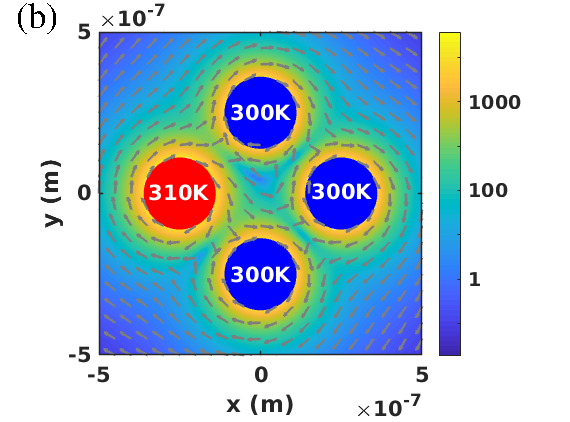}
	\caption{(a) Plot of the transmission coefficients $\mathcal{T}_{12}$, $\mathcal{T}_{13}$ and $\mathcal{T}_{31}$ for the configuration in Fig.~\ref{Fig:Sketch} with $d = 300\,{\rm nm}$ and radius $r = 100\,{\rm nm}$ for four nanoparticles made of a WSM. The three dipolar resonances are at $\omega_{m = +1} = 1.486\times10^{13}\,{\rm rad/s}$, $\omega_{m = 0} = 7.714\times10^{13}\,{\rm rad/s}$, and $\omega_{m = -1} = 3.856\times10^{14}\,{\rm rad/s}$. The inset shows the ratio $\mathcal{T}_{31}/\mathcal{T}_{13}$ illustrating the asymmetry of clockwise and counterclockise heat flux due to the non-reciprocity.
		 (b) Normalized mean Poynting vector (arrows) and its magnitude (colour scale) for the Hall configuration in Fig.~\ref{Fig:Sketch} with WSM nanoparticles having the temperatures as indicated in the plot. For the calculation of the mean Poynting vector we have used the method described in Ref.~\cite{Ott_JPE}. }
  \label{Fig:TransmissionWeyl}
\end{figure}

As a consequence of the non-reciprocity, even for four identical nano-particles we have $\mathcal{T}_{ij} \neq \mathcal{T}_{ji}$.  This asymmetry is a key condition for the existence of an ATHE. Note that this condition is not restricted to dipolar systems. For bigger particles the higher order modes (multipoles) should be taken into account to derive the optical responses (generalized susceptibility) of particles and transmission coefficients.
As can be seen in the inset of Fig.~\ref{Fig:TransmissionWeyl}(a), in the dipolar case, the non-reciprocity is for the used parameter set especially large for the $m = -1$ resonance. For other choices of parameters also the $m = +1$ resonance can significantly contribute to the non-reciprocity. When  $\mathcal{T}_{ij} \neq \mathcal{T}_{ji}$ we also have $\varphi_{ij} \neq \varphi_{ji}$. In particular, in the $C_4$ symmetric configuration shown Fig.~\ref{Fig:Sketch} we find 
\begin{equation} 
	\varphi_{13} = \varphi_{32} = \varphi_{24} = \varphi_{41} \neq \varphi_{14} = \varphi_{42} = \varphi_{23} = \varphi_{31}.
	\label{Eq:symmetry}
\end{equation}
In other words, due to the non-reciprocity the heat flow clockwise and counterclockwise is different as observed for the persistent heat current~\cite{zhufan,zhufan2, Ott_JPE} leading to a circular heat flux~\cite{OttEtAl2018} which is at the heart of the thermal Hall effect~\cite{PBA_PRL2016,Ott_JPE}. The circular heat flux around the WSM particles is shown in Fig.~\ref{Fig:TransmissionWeyl}(b).

Now let us assume that the two particles along the $\mathbf{x}$-axis are connected to two thermostats fixing their temperatures so that a heat flux flows through the system between these two particles. While in a reciprocal system, no heat flux exists between the two other unthermostated particles  for symmetry reasons, in non-reciprocal systems a Hall flux appears giving rise to a transversal temperature gradient in steady state. The magnitude of this Hall effect can be evaluated using the relative Hall temperature difference
\begin{equation}
	R_H = \frac{T_3-T_4}{T_1-T_2} \label{Eq:relative1}
\end{equation}
where all temperatures are the steady states temperatures. In linear response regime, this expression can also be rewritten in term of the thermal conductance 
\begin{equation}
	G_{ij} =  3\int_0^\infty\!\!\frac{\rd \omega}{2 \pi}\, \frac{\partial \Theta (\omega,T)}{\partial T} \biggr|_{T = T_j} \mathcal{T}_{ij}(\omega) 
\end{equation}
between the  $i^{th}$ and the $j^{th}$ particle as~\cite{PBA_PRL2016}
\begin{equation}
	R_H = \frac{G_{13}G_{24}-G_{14}G_{23}}{\underset{j\neq 3}{\sum}G_{j3}\underset{j\neq 4}{\sum}G_{j4}-G_{43}G_{34}}
	\label{Eq:relative2}.
\end{equation}
Taking the symmetries in Eq.~(\ref{Eq:symmetry}) due to the C$_4$ configuration into account we can rewrite $R_H$ as
\begin{equation}
	R_H = \frac{G_{13} - G_{31}}{G_{13} + G_{31} + 2 G_{34}}
	\label{Eq:relative3}.
\end{equation}
This expression clearly shows that the assymetry in clockwise and conter-clockwise heat flow expressed by $G_{13} - G_{31}$ introduced by the non-reciprocal Weyl metal is at the heart of the ATHE. Alternatively, to determine $R_H$  the temperatures $T_1$ and $T_2$ are fixed and the equilibrium temperatures of particles $3$ and $4$  are calculated  numerically  by solving the system of overall power flow ~\cite{Ott_JPE}
\begin{eqnarray}
\phi_{3}(T_3,T_4)=\underset{j}{\sum}\varphi_{j3}=0,\\
\phi_{4}(T_3,T_4)=\underset{j}{\sum}\varphi_{j4}=0
\label{Eq:temp_eq}.
\end{eqnarray}
These two methods  are equivalent and give the same results for $R_H$ provided that $|T_1 - T_2| \ll \min(T_1,T_2)$.
\begin{figure}
  \centering
  \includegraphics[angle=0,scale=0.45,angle=0]{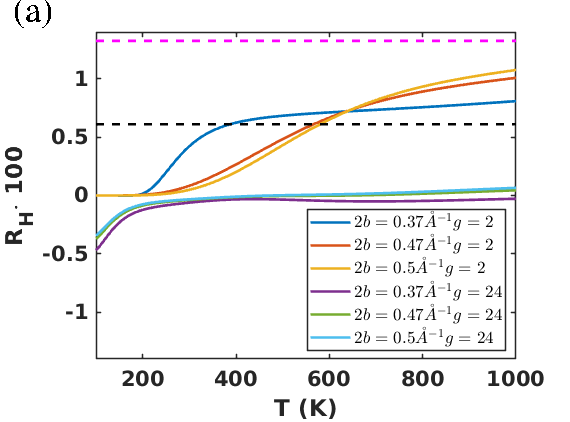}
  \includegraphics[angle=0,scale=0.45,angle=0]{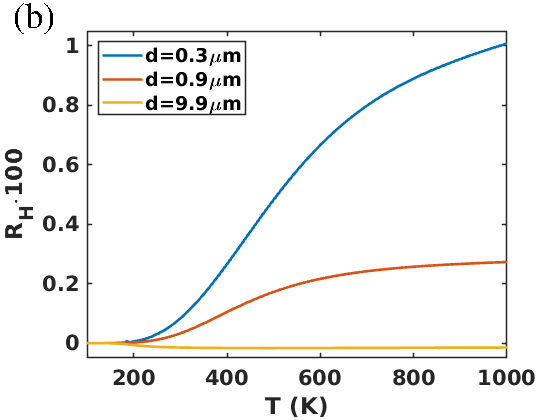}
	\caption{($R_H$ from Eq.~(\ref{Eq:relative3}) as function of temperature. (a) $R_H$ for different WSM parameters $g = 2, 24$ and $2 b = 0.37 \mathring{A}^{-1},  0.47 \mathring{A}^{-1}, 0.5 \mathring{A}^{-1}$ with $r = 100\,{\rm nm}$ and $d = 300\,{\rm nm}$. The horizontal dashed black (pink) line represents the value of $R_H$ for InSb with a magnetic field of $0.1\,{\rm T}$  ($1\,{\rm T}$) in z-direction and $T = 300\,{\rm K}$. (b) $R_H$ for different distances using $g = 2$ and $2 b =  0.47 \mathring{A}^{-1}$.}
  \label{Fig:RH}
\end{figure}

In Fig.~\ref{Fig:RH}(a) we show the temperature dependence of the relative Hall temperature difference $R_H$ for different WSM parameters. It is apparent that the magnitude and the directionality highly depend on these parameters. For a number of Weyl points $g = 2$ we find that $R_H$ is mainly positive, i.e. particle 3 is heated up more efficiently than particle 4. This tendency could be anticipated from Fig.~\ref{Fig:TransmissionWeyl}(a) which clearly shows that $ \mathcal{T}_{13} < \mathcal{T}_{31}$ and from Fig.~\ref{Fig:TransmissionWeyl}(b) where the Poynting vector in the inner circle of the particles is bend towards particle 3. Furthermore, the ATHE becomes stronger for higher temperatures or smaller values of momentum-separation $2b$ of the Weyl nodes. The magnitude of Hall effect is comparable to the case of InSb particles with $B = 0.1-1\,{\rm T}$. On the other hand,  with $g = 24$ the Hall effect is the strongest for low temperatures and the direction of Hall flux is reversed, i.e.\ particle 4 is heated up more efficiently than particle 3. This can be traced back to the fact that for $g = 24$ the dipole resonance at $\omega_{m = +1}$ dominates the heat flux so that the circularity is inverted and $ \mathcal{T}_{13} > \mathcal{T}_{31}$. Finally, in Fig.~\ref{Fig:RH} it can be seen that the strength of the ATHE becomes smaller in the far-field regime and even changes its directionality as also observed for the circular heat flux~\cite{Ott_JPE}. 

In conclusion, we have demonstrated that the optical non-reciprocity in WSM nanoparticles networks induces circular heat flux around the particles resulting in an ATHE without applied external magnetic field. We have shown that the intrinsic time-reversal symmetry breaking in these systems gives rise to a Hall effect  which is comparable to those observed in magneto-optical networks with magnetic fields of $B = 0.1$ to $1\,{\rm T}$. It is worthwile to note that the non-reciprocity in WSM should also be responsible of  persistent heat flux, persistent angular momentum and spin~\cite{OttEtAl2018,zhufan,zhufan2,Silveirinha,Zubin2019} of thermal radiation. Besides its fundamental interest the ATHE for WSM opens up the possibility to  control the direcionality of the radiative heat flux in nanoscale systems without the necessity of applying strong magnetic fields. This opens the route for an alternative way for thermal management and heat flux guiding in nanoscale systems.

\begin{acknowledgements}
S.-A.\ B. acknowledges support from Heisenberg Programme of the Deutsche Forschungsgemeinschaft (DFG, German Research Foundation) under the project No. 404073166 and discussions with R. Messina. 
\end{acknowledgements}

\end{document}